\documentclass[aps,prl,showpacs,floatfix,twocolumn,superscriptaddress,
amsmath,amssymb]{revtex4}

\usepackage{graphicx}

\begin{document}

\title{Chaos and rectification of electromagnetic wave in a lateral
semiconductor superlattice}
\author{Kirill N. Alekseev}
\author{Pekka Pietil\"{a}inen}
\affiliation{Department of Physical Sciences,
P.O. Box 3000, FIN-90014 University of Oulu, Finland}
\author{Alexander A. Zharov}
\affiliation{Institute of Microstructure Physics, Russian Academy of Sciences,
Nizhnii Novgorod 603950, Russia}
\author{Feodor V. Kusmartsev}
\affiliation{Department of Physics,
Loughborough University, Loughborough LE11 3TU, United Kingdom}

\begin{abstract}
We find the conditions for a rectification of electromagnetic wave in a lateral
semiconductor superlattice with a high mobility of electrons. The rectification
is assisted by a transition to a dissipative chaos at a very high mobility.
We show that mechanism responsible for the rectification is a creation of
warm electrons in the superlattice miniband caused by an interplay of the effects 
of nonlinearity and finite band width.
\end{abstract}
\maketitle

The research of nonlinear transport properties of the strongly coupled
semiconductor superlattices (SLs) under the action of constant and
alternating fields is the subject of traditional interest associated with the
great potential of numerous
applications in microwave and terahertz technologies \cite{rev-ssl}.
In recent years the primary goal has been investigations of novel
{\it strongly nonlinear phenomena} at the miniband transport in SLs, such as
dissipative chaos \cite{diss-chaos} and
excitation of Bloch oscillations under the action of a pure ac
electric field \cite{sb-first,sb-other}. The last effect can be
considered as a new mechanism of the rectification of THz radiation;
the generated dc voltage is either integer \cite{sb-first} or fractionally
quantized \cite{sb-other}.
\par
The recent experimental progress in the observation of the
influence of intensive THz electromagnetic radiation on voltage-current
characteristics of SLs \cite{exp-thz} has stimulated advances in the
theory \cite{diss-chaos,sb-first,sb-other}. Observations of
strongly nonlinear properties of the miniband transport in SLs subjected to an
intensive electromagnetic field require
a relatively high level of the carrier density together with the high 
carrier mobility \cite{diss-chaos,sb-first,sb-other}.
However, due to the experimental conditions associated with bulk SLs
it is difficult to meet  simultaneously these two requirements, because of a
relatively low electron mobility and a formation of high field domains inside
heavy doped SLs.
\par
These limitations can be overcome in the lateral semiconductor superlattices
(LSSLs) \cite{sakaki}, and especially in those LSSLs which are
fabricated using the cleaved edge overgrowth (CEO) technique \cite{exp-samples}.
In these LSSL devices a two-dimensional electron gas resides in an atomically
precise one-dimensional superlattice potential. CEO LSSLs
are most suitable for an observation of Bloch oscillations: they have high
electron mobility  and can deliver high currents without formation of high
field and low field domains \cite{exp-samples}.
In these respects the LSSLs are promising objects for investigations
of various strongly nonlinear phenomena.
\par
In this Letter, we describe a novel mechanism of rectification of
electromagnetic radiation in a superlattice that is different from the
mechanism of AC-induced Bloch oscillations \cite{sb-first} and that always
results in a generation of an unquantized bias.
We consider an interaction of a plane electromagnetic wave
with 2D electrons in a single miniband of LSSL, and incorporate
the feedback effects of electron motion on the incident field.
These feedback effects are responsible for the nonlinearity of the problem.
We show that for some field strengths and frequencies, the interplay of effects
of the nonlinearity and the finite miniband width causes a weak asymmetry
of a distribution function of electrons in the momentum space with respect
to the center of Brillouin zone. This deformation of nonequilibrium
distribution function creates a direct current that is the reason of
an appearance of bias in the wave transmitted through the LSSL.
In mathematical description, this effect  corresponds to the
symmetry-breaking bifurcation in nonlinear dynamical systems \cite{sb-def}.
We also show that the interaction of the electromagnetic radiation with
miniband electrons of the very high mobility can result in a dissipative
chaos instead of a rectification.
\par
Consider a plane electromagnetic wave $E_{in}=E_0\cos(\Omega t- k z)$, which
is polarized along the superlattice axis, being incident normally to the
surface of LSSL located at $z=0$. The LSSL sits on a semi-insulating background
with an average dielectric constant $\epsilon_0$.
We assume that both the superlattice thickness in the $z$-direction and the
superlattice length along its axis are much less than the
characteristic scale of the electromagnetic wave in the medium, $2\pi/k$.
To study the electron transport through a single miniband, a spatially
homogeneous LSSL with the period $a$ and the miniband width $\Delta$,
we use the tight-binding energy-quasimomentum
dispersion relation 
$\varepsilon(k_{el})=(\Delta/2)\left[1-\cos(k_{el} a)\right]$
($k_{el}$ is the electron wave vector along the superlattice axis).
The dynamics of miniband electrons is described by the following nonlinear
balance equations \cite{dodin98}
\begin{subequations}
\label{balance0}
\begin{eqnarray}
\dot{v}&=&-U w -\gamma v,\label{balance0_a}\\
\dot{w}&=&U v -\gamma (w-w_{eq}),\label{balance0_b}\\
U&=&U_{in}(t)-\kappa v,\label{balance0_c}
\end{eqnarray}
\end{subequations}
where $U_{in}(t)=(2 e a/\hbar(1+\epsilon_0^{1/2})) E_{in}(t,z=0)$ and
$U(t)=(e a/\hbar) E_{out}(t,z=0)$
($E_{in}(t,z)$ and $E_{out}(t,z)$ are the fields of the incident wave and
the wave transmitted by LSSL, respectively).
The relaxation processes for miniband electrons are characterized by
a scattering constant $\gamma$.
The electron variables $v=m_0 \overline{V} a/\hbar$,
$w=(\overline{\varepsilon}-\Delta/2)(\Delta/2)^{-1}$ and $w_{eq}$
are the scaled electron velocity, the scaled
electron energy, and the equilibrium value of the scaled electron energy,
respectively, and $m_0=(2\hbar^2)/(\Delta a^2)$ is the effective mass
at the bottom of the miniband. The lower (upper) edge
of the miniband corresponds to $w=-1$ ($w=+1$), and the value of
$w_{eq}$ $(w_{eq}<0)$ is a function of lattice temperature.
The scaled variables $v(t)$ and $w(t)$ are proportional to
the variables $\overline{V}(t)$ and $\overline{\varepsilon}(t)$,
which are the electron velocity and  the energy averaged over the 
time-dependent
distribution function satisfying the Boltzmann equation.
\par
The first two Eqs. of set (\ref{balance0}) are well-known superlattice balance
equations \cite{ignatov76}; their solution for a given field $U(t)$
is exactly the same as the solution of the time-dependent Boltzmann 
equation for a tight-binding lattice.
Eq.~(\ref{balance0_a}) describes the acceleration of
electrons under the action of the electric field $U(t)$ and their slowing
down caused by an effective friction due to the scattering, while the second
Eq.~(\ref{balance0_b})
describes the balance of electron's energy gain under the action of
the electric field and the energy loss due to scattering.
These equations are valid if the standard conditions of semiclassical approach,
$e E_0 a<\Delta$, $\hbar\omega<\Delta_g$, $e E_0 a<\Delta_g$ ($\Delta_g$ is the width of
minigap) \cite{rev-ssl}, are satisfied.
\par
The Eq.~(\ref{balance0_c}) is derived by modeling
the LSSL as an equivalent current screen of infinitesimal thickness
with the use of the Maxwell equations and appropriate boundary conditions
\cite{dodin98,rupasov}.
Such approach is valid if superlattice thickness is
much less than the wavelength $2\pi/k$ \cite{dodin98}; this condition is
easily satisfied for typical LSSL irradiated by submillimeter or millimeter
electromagnetic waves.
This Eq.~(\ref{balance0_c}) describes the back influence of the
electron current in LSSL on the dynamics of miniband electrons making
the whole set of Eqs.~(\ref{balance0}) nonlinear.
The degree of nonlinearity is controlled by
the value of parameter $\kappa=(4\pi e^2 N_s)/m_0 c
(1+\epsilon_0^{1/2})$,
where $c$ is the speed of light and $N_s$ is the  areal electron density.
\par
The dynamics of electrons in LSSL as well as the time-dependence of the
transmitted and reflected waves is determined by an interplay of the 
relaxation ($\propto\gamma$) and the nonlinearity ($\propto \kappa \sim  N_s$).
Therefore, it is convenient to introduce a
single parameter, $\Gamma=\left( \gamma/\kappa\right)^{1/2}$, controlling
the relative contribution of these two basic processes. For this we introduce
the scaled time $\tau=\omega_0 t$, $\omega_0=\left( \kappa\gamma\right)^{1/2}$,
scaled fields $u=U/\omega_0$, $u_{in}=U_{in}/\omega_0$,
and rewrite Eqs. (\ref{balance0}) as
\begin{eqnarray}
\dot{v}&=&-\left( u_{in}(\tau)-\Gamma^{-1} v\right) w -\Gamma v,\nonumber\\
\dot{w}&=&\left( u_{in}(\tau)-\Gamma^{-1} v\right) v -\Gamma (w-w_{eq}),
\label{balance1}
\end{eqnarray}
\begin{figure}[htbp!]
\includegraphics[width=1\linewidth]{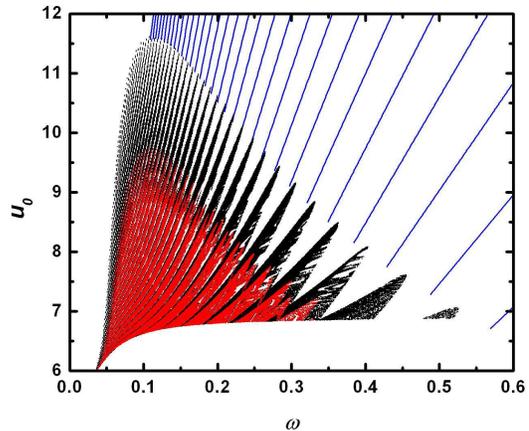}
\caption{(color).
The regions of the rectification (black) and the chaos (red) in the 
frequency -- field
strength plane for $\Gamma=0.1$. Straight blue lines mark roots of Bessel
function.}
\label{figps1}
\end{figure}
\begin{figure}[t!]
\includegraphics[width=0.8\linewidth]{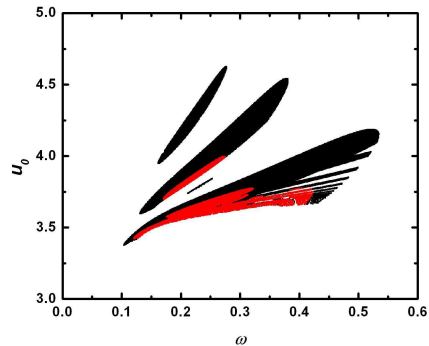}
\caption{(color).
Same as in Fig.\ref{figps1}, but for $\Gamma=0.2$.}
\label{figps2}
\end{figure}
where the overdot means the differentiation with respect to the value $\tau$,
$u_{in}(\tau)=u_0\cos\omega\tau$ with $\omega=\Omega/\omega_0$ and
$u_0= (2 e a E_0)/\hbar\omega_0 (1+\epsilon_0^{1/2})$.
The wave transmitted by LSSL is
$u(\tau)=-\Gamma^{-1} v+u_0\cos\omega\tau$.
\par
\begin{figure}[htbp!]
\includegraphics[width=0.7\linewidth]{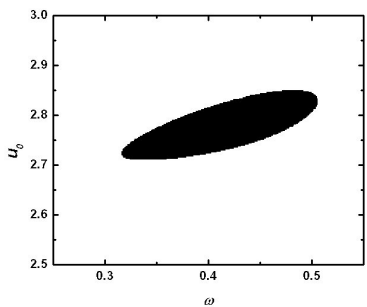}
\caption{
The region of  the rectification in the frequency -- field
strength plane for $\Gamma=0.3$.}
\label{figps3}
\end{figure}
\begin{figure}[htbp!]
\includegraphics[width=0.8\linewidth]{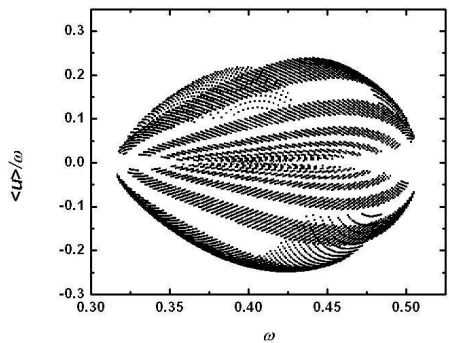}
\caption{
The dependence of the spontaneously generated dc bias $\langle 
u\rangle/\omega$ on
frequency $\omega$ for $\Gamma=0.3$.}
\label{figps4}
\end{figure}

We have performed a systematic numerical search for stationary solutions
($t\gg\Gamma^{-1}$) of Eqs.~(\ref{balance1}) which
have a nonzero dc component of the outgoing wave, $\langle u\rangle\neq 0$
(the averaging is made over a period of the ingoing wave $T=2\pi/\omega$).
We also detect the chaotic attractors characterized by a
positive value of the maximal Lyapunov exponent \cite{diss-chaos}.
In contrast, the Lyapunov exponent is always zero for the limit cycles.
The locations of chaotic attractors and symmetry-broken periodic attractors
with $|\langle u\rangle|\geq 10^{-4}$ in the $\omega u_0$-plane for different
values of $\Gamma$ are shown in Figs.~\ref{figps1}-\ref{figps3}.
The structure of the $\omega u_0$-plot shown in Fig.~\ref{figps1} for
$\Gamma=0.1$ is quite typical for the behavior of our system at weak
damping. It consists of the stripes for solutions supporting rectification
and some area of chaotic solutions located at the lower values of $u_0$.
For a weak damping and a low frequency, ``symmetry-broken stripes'' are located
near the lines of zero order Bessel function roots defined as
$J_0 \left( u_0/\omega\right)=0$
(Fig.~\ref{figps1}, only about 30 roots are shown).
With an increase of damping ($\Gamma=0.2$, Fig.\ref{figps2}), the
region for an existence of chaotic attractors in the $\omega 
u_0$-plane shrinks,
while the regions of nonchaotic attractors with $\langle u\rangle\neq 0$ occupy
a few wide stripes (cf. Figs.\ref{figps1} and \ref{figps2}).
Finally, with a further increase of $\Gamma$, chaos completely disappears and
attractors with $\langle u\rangle\neq 0$ are located in a single compact
region in the $\omega u_0$-plane (Fig.\ref{figps3}). For $\Gamma\geq\Gamma_{cr}
\approx 0.4$ we didn't observe the effect of rectification anymore.
The generated dc voltage is unquantized (Fig.\ref{figps4}).
This is in contrast to the case of rectification making use of the mechanism of
ac-induced Bloch oscillations, where the ratio
$\langle u\rangle/\omega$ can be either an integer or a fractional number
for a weak damping \cite{sb-first,sb-other}.
The efficiency of transformation of ac field to dc field,
$\langle u\rangle/u_0$, is rather low; in conditions of Fig.\ref{figps3}
it is less than four percents overall.
\par
Physical mechanism responsible for the rectification of electromagnetic wave
in LSSL will be our focus in the remainder of this paper.
There are three main distinct physical factors which can contribute to the
generation of unquantized dc voltage in our problem: finite band width,
nonlinearity, and dissipation. In order to distinguish an influence of these
factors, we consider the limit $\Gamma\rightarrow 0$ in the
Eqs. (\ref{balance1}) and get
\begin{equation}
\label{theta-eq}
\dot{\theta}+\Gamma^{-1}\sin\theta = -u_{0}\cos\omega\tau,
\end{equation}
where $v=-\sin\theta$, $w=-\cos\theta$ and $u=-\dot{\theta}$.
Eq. (\ref{theta-eq}) describes a time evolution of electrons within
the first Brillouin zone ($|\theta|<\pi$)
for a negligible scattering but with an account of nonlinearity.
Unbiased AC field can not create {\it rotational states} with
$\langle\dot{\theta}\rangle\neq 0$ in the overdamped pendulum,
therefore the AC-induced Bloch oscillations \cite{sb-first}
can not exist here. However, the ac field can excite
symmetry-broken swinging states with
$\theta_0\equiv\langle\theta\rangle\neq 0$ \cite{sb-def}.
In a small vicinity of the
symmetry-breaking bifurcation, a stationary solution of  Eq. (\ref{theta-eq})
can be presented in the form \cite{miles}
$\theta=\theta_0+\alpha\sin(\omega\tau+\mu)$, where $\alpha$ and $\mu$ are
constants dependent on the parameters of LSSL. Substituting this expression in
Eq. (\ref{theta-eq}) and equating zero harmonics, we have
\begin{equation}
\label{alpha-cond}
J_0(\alpha)\sin\theta_0=0.
\end{equation}
Eq. (\ref{alpha-cond}) shows that either $\theta_0=0$ (a symmetric solution) or
$\theta_0\neq 0$, but  $\alpha$ should satisfy $J_0(\alpha)=0$
(a boundary of symmetry-breaking, {\it cf.} ref. \cite{miles}).
Analogously, equating first harmonics, we can find the dependence of amplitude
$\alpha$ on $u_{0}$ and $\omega$ for a symmetric solution,
as well as the dependence $\theta_0$ on $u_{0}$ and $\omega$ at the boundary of
symmetry-breaking \cite{kna-unbubl}. This analysis shows that the behavior of
symmetric and symmetry-broken solutions is quite different:
For a symmetric solution $\alpha$ increases with increase of $u_{0}$, while for
an asymmetric regime $\theta_0$ increases but $\alpha$ stays constant.
\par
Therefore, we should distinguish two regimes of the wave transmission through
a LSSL.
In the first regime, the interaction of ac field with an electron gas can led
only to a difference between the amplitudes of incident, $u_{0}$, and
transmitted, $u_{max}=\omega\alpha$, waves due to a nonlinear dependence of
$\alpha$ on $u_{0}$ ({\it cf.} \cite{dodin98}).
The value of $\theta_0$ is zero indicating that the distribution function of
electrons in $k_{el}$-space is symmetric in respect to the center of
Brillouin zone. With an increase of $u_{0}$ this regime continues
until $\alpha$ reaches the value corresponding to one of
the Bessel roots.
In this second regime of a strong interaction between electrons and
radiation, a variation of amplitude of the incident wave practically doesn't
result in a change of amplitude of the transmitted wave: $u_{max}\propto
\alpha$=const independently on $u_0$, at least near the boundary of such
regime. Instead, the energy of absorbed radiation goes to an effective heating
of electron gas and a creation of ``warm electrons''. The distribution function
of these ``warm electrons'' is slightly asymmetric in respect to
the center of Brillouin zone ($\theta_0\neq 0$). {\it In the presence
of scattering the asymmetry of distribution function results in a directed
current and in an appearance of dc component of the outgoing wave.}
\par
Consider the physics of a rectification of the wave with
a low frequency, $\omega\ll 1$, and a large amplitude, $u_0\agt\Gamma^{-1}$,
in more detail.
In this case our analysis shows \cite{kna-unbubl} that
$\alpha\rightarrow u_0/\omega$ and therefore the condition for
symmetry-breaking bifurcation in Eq.~(\ref{theta-eq}) becomes
$J_0 \left( u_0/\omega\right)=0$ (we also have checked it solving
Eq.~(\ref{theta-eq}) numerically).
On other hand, this condition is in a reasonable agreement with the
results of numerical
simulations of rectification at small  $\Gamma$ employing
balance equations (\ref{balance1}) (see Fig.\ref{figps1} for $\Gamma=0.1$).
Further, we find that the large amplitude of ac field is almost unchanged after
passing trough LSSL, $u_{max}=\omega\alpha\approx\omega\left( u_0/\omega\right)
=u_0$. The absolute value of generated dc bias is very small:
$|\langle u\rangle|\simeq 10^{-4}$ for $\Gamma=0.1$ (however, such values are
still quite distinguishable numerically from artificial dc effects arising
due to round off errors that result in $|\langle u\rangle|\alt 10^{-8}$).
The expression $J_0 \left( u_0/\omega\right)=0$ determines
{\it the condition for dynamic localization} of miniband electrons
driven by {\it a given ac field} \cite{localization}.
Both quantum mechanical \cite{hone} and semiclassical \cite{romanov-ftt01}
considerations of the dynamic localization show that a forward motion of an 
electron is transformed to its local oscillations.
These localized electrons strongly interact with radiation.
The localization of every electron under the action of a strong ac field is
a coherent control effect \cite{localization}. However, different oscillating
electrons have different and random phases in the presence of scattering; 
interaction of many electrons with the radiation results in an effective 
electron's heating \cite{romanov-ftt01}. It was shown recently \cite{cannon}
that the overheating may lead to the symmetry-breaking instability resulting 
in dc. The development of this instability and its next stabilization
is a rather complicated process which is associated with a nonlinear
feedback action of electrons on the incident field
(the terms $+\Gamma^{-1} v w$  and $-\Gamma^{-1} v^2$ in Eqs.~(\ref{balance1})).
Importantly, the nonlinear effects can not be ignored anymore in the 
conditions of localization even for a low electron density $N_s$.
Thus, when a dissipation is weak and the self-consistent field
inside LSSL is only slightly different from the incident ac field,
just an interplay of the effects of dynamic localization and nonlinearity
provides a conversion of radiation into a very small dc bias.
\par
For available LSSLs fabricated with the use of CEO technique,
$a=15$ nm, $\Delta=12$ meV, $\Delta_g=85$ meV, $\epsilon_0=13$ (GaAs),
$N_s=6\times 10^{11}$ cm$^{-2}$ and $\gamma^{-1}=10$ ps at temperature $0.3$ K
\cite{exp-samples}, we estimate $\Gamma=0.6>\Gamma_{cr}\approx 0.4$.
Because $\Gamma\propto (N_s\tau\Delta a^2)^{-1/2}$, a further increase of
the electron's density, the mobility, as well as an increase of the miniband
width should result in a possibility to observe the
predicted effects of rectification and chaos.
In physical units, the desirable field strength
$u^{(sb)}_0\approx 2.75$ and the frequency $\omega\approx 0.35$
(cf. Fig.~\ref{figps3}) correspond to $E^{(sb)}_0\approx 825$ V/cm and
$\Omega/2\pi\approx 9$ GHz,
that can be easily  reached within the range for existing Gunn diodes.
At the same time, the conditions of applicability
of the semiclassical superlattice equations are well satisfied for such values
of parameters (in particular,
$e a E_0/\Delta\approx u_0 (\hbar\omega_0/\Delta)\simeq 10^{-2}\ll 1$).
\par
In conclusion, we should notice that likely similar effects of a
resonant photovoltaic response of 2D electron gas in semiconductor
heterojunctions to microwave
and far-infrared radiations have been observed in the experiments
\cite{photovolt-exp1,photovolt-exp2,photovolt-exp3}.
These photovoltaic effects were attributed in Refs.
\cite{photovolt-exp1,photovolt-exp2,photovolt-exp3} to a ``resonant heating''
of intraband electrons confined by the different methods, such as an
excitation of the discrete 2D plasmons
\cite{photovolt-exp1}, an application of the quantized magnetic field
\cite{photovolt-exp2} or a creation of the
field-effect confined quantum dots \cite{photovolt-exp3}.
In contrast, we consider the localization of electrons in the superlattice
miniband induced by the ac electric field itself (dynamic localization).
Possible realizations of the symmetry-breaking bifurcations
in the particular experimental situations
\cite{photovolt-exp1,photovolt-exp2,photovolt-exp3}
must be further investigated.
\par
This research was supported by Academy of Finland, Royal Society, Humboldt
Foundation and RFBR.

\end{document}